\documentstyle[psfig,epsf,aps,12pt]{revtex}

\begin{document}
\draft

\title{Noise Sustained Propagation: local vs. global noise}

\author{
M. L\"ocher$^{\ddagger}$\footnote{Now at: Siemens Corporate Research, 755 College Rd., Princeton, NJ 08550}
, N. Chatterjee$^{\#}$, 
F. Marchesoni$^{\star}$, 
W. L. Ditto$^{\ddagger}$ and E. R. Hunt$^{\#}$\\ 
$^{\ddagger}${\it School of Physics, Georgia Institute of Technology,
 Atlanta, GA  30332-0430}\\
$^{\#}${\it Dept. of Physics and Astronomy, Ohio University, Athens, OH 45701\\}
$^{\star}${\it Instituto di Fisica della Materia, Universita' di Camerino,
I-62032 Camerino (Italy), and\\
 Dept. of Physics, University of Michigan, Ann Arbor, MI 48109-1120.} 
}

\date{\today}
\maketitle

\begin{abstract} 
We expand on prior results on noise supported signal propagation in 
arrays of coupled bistable elements.
We present and compare experimental and numerical results for kink propagation 
under the influence of local and global fluctuations.
As demonstrated previously for local noise, an optimum range of 
global noise power exists for which the medium 
acts as a reliable transmission ``channel''.
We discuss implications for propagation failure in a 
model of cardiac tissue and present a general theoretical framework 
based on discrete kink statistics.
Valid for generic bistable chains, the theory captures the essential features observed in 
our experiments and numerical simulations.

\end{abstract}

%\narrowtext

\section{Introduction}
Information transfer through nonlinear systems in the presence of fluctuations 
has been extensively studied in 
the context of stochastic resonance (SR) \cite{sr}. The frontier of this research 
has shifted towards systems 
with spatial degrees of 
freedom over the 
past few years. While the efforts initially were directed towards enhancing 
the basic SR effect \cite{SPT_SR}, 
recent work has demonstrated that noise can also 
sustain wave propagation \cite{NEP1,NEP2,NEP_SOC}. 
In early studies, Jung {\it et al} \cite{NEP1} showed that noise can sustain 
spiral waves in a caricature model 
of excitable media. The assisting role of noise for one- and two-dimensional, 
nonlinear wave propagation was 
experimentally confirmed 
by K\'ad'ar {et al.} \cite{NEP1} in a chemical medium and by L\"ocher {\it et al.} \cite{NEP2} in an array of 
coupled electronic resonators.
Noise enhanced propagation (NEP) for {\it periodic} signals was explored by Lindner {\it et al.} \cite{NEP1} 
in a chain of coupled, overdamped bistable oscillators. 
The authors of Ref. \cite{NEP_SOC} furnish evidence for self-organized criticality 
underlying the creation 
and propagation of waves by noise in a chemical subexcitable medium.
So far, all experiments and simulations on NEP utilized {\it local and additive} noise.
To our knowledge, no comparative studies on the effectiveness of local vs. global, 
additive vs. multiplicative noise have been attempted.

In this paper, we investigate experimentally the effectiveness of global 
noise as compared to local noise for the propagation of a signal in 
a chain of coupled bistable elements. 
Expanding on earlier reports \cite{NEP2}, the experiments are performed using a 
16 x 16 array of diode resonators driven in the stable period-2 regime. 
A bias consisting of a second drive at half the main frequency 
renders one phase more stable and a phase kink can be made to propagate 
across the array. 
For an intermediate value of the coupling resistors and small bias, 
we observe propagation failure. Adding noise to the drive of each element, 
either from a single source or from individual sources - corresponding to 
global and local noise, respectively - greatly increases the 
chances of successful signal transmission.
Digital simulations of NEP in a simple model of cardiac tissue compare 
favorably with these observations and allow us to investigate the 
effects of parameter mismatch between the elements.
%Finally, we present preliminary insights into the unique role 
%multiplicative noise plays for noise sustained kink motion.
At last, we present theoretical insights borrowed from discrete kink statistics 
which describe the phenomena in reasonable agreement.
%preliminary insights into the unique role multiplicative noise plays for 
%noise sustained kink motion.

%%%%%%%%%%%%%%%%%%%%%%%%%%%%%%%%%%%%%%%%%%%%%%%%%%%%%%%%%%%%%%%%%%%%%%%%%%%%%%%%%%%%%%%%%%%%%%%%%%%%%%%%%
\section{Experimental Results}
\label{ExpResults}
The experimental setup consists of 256 coupled diode resonators, of which 
one element is shown in Fig.\ref{circuit}. The elements are arranged
in a $16 \times 16$ array with periodic boundary conditions connected along two opposite edges. 
Each diode resonator works as a bistable element when driven
in its period-2 state. We break the phase symmetry by adding to the drive a second 
sinusoidal signal at half its frequency. 
This bias renders one phase more stable.  We refer to the less stable phase as the 
metastable state. By inducing a phase change at one 
edge of the array a one-dimensional wavefront comprised of phase kinks will 
travel towards the detector at the other edge.  The noise generators were 
constructed using the shot noise generated by a current through a {\it pn} 
junction diode as a source.

It is clear that in the absence of noise, a local phase jump will lead to 
a ``domino effect'' only
if the bias and coupling are strong enough.  The energetically lower 
phase then propagates into the metastable
phase in the form of a moving kink.  For identical elements, 
the speed of this moving interface depends on both the coupling 
strength and the amplitude of the applied bias.  If the latter 
two parameters are chosen low enough, kinks in discrete systems will fail to propagate.
In the experiment, there is a third factor contributing to kink 
trapping, namely heterogeneity of the chain. In our system, the variation of key parameters of 
the diode resonators and the difference in local noise power is as high as 
10 \%. Arranging the array with periodic boundary conditions in one dimension
and preserving the motion of wavefronts in the other dimension, we considerably
reduce inhomogenities by effectively averaging over 16 elements. 

Figure. \ref{ExpKinkVel} displays the 
 measured kink velocities in the absence of noise as a function of bias for an 
intermediate value of the coupling resistors.  
The velocity decreases approximately linearly with the bias down to a cutoff value of 0.9 units where it rapidly falls off to zero.
We operate the array at a bias of 0.6 units for which the system is not capable of deterministic kink motion.

Our experimental results are given in terms of the arrival-probability curves shown in Fig. \ref{ExpRise}. The procedure for obtaining these are analogous to those described in \cite{NEP2}:
We reset all resonators to be in the metastable state and then induce a 
phase flip at one edge of the array. Any successful kink propagation generates a step function at the detector, which we average over approximately 100 events, resulting in a smooth rise curve 
(solid lines).
In order to quantify transmission degeneration by noise nucleated spurious 
signals we repeat the same experiment without inducing a phase flip 
initially (dashed lines). 

The arrival-probability curves for both local and global noise at
five different noise strengths are shown in Fig. \ref{ExpRise}.
For local, i.e. independent from site to site, noise we obtain results 
analogous to those previously reported \cite{NEP2}. For a local noise background of 
less than 0.0025 $mV^2/Hz$ (Fig. \ref{ExpRise}(a)) kinks remained trapped longer than 
the measurement time of 5000 drive cycles. At 0.0121 $mV^2/Hz$ the 
signals arrive with a wide distribution of travel times and a slow mean 
(Fig. \ref{ExpRise}(b)). The arrival times become shorter with
increasing noise strengths (Figs. \ref{ExpRise}(c) and \ref{ExpRise}(d)). 
Finally, at 0.0625 $mV^2/Hz$ a substantial number of false starts 
(dashed lines) corrupt the detection of the original input signal.

We found that the qualitative behavior of the chain under the influence of 
global noise is similar, but the onset of detectable kink propagation 
is found at much lower noise strengths. We observe no kink propagation 
below noise levels of 0.001 $mV^2/Hz$.
Slow and disperse kink motion occurs for noise 
levels of 0.0025 $mV^2/Hz$ (Fig. \ref{ExpRise}(a)). Higher average kink speeds and 
less fluctuations are recorded for levels of 0.0121 and 0.0256 $mV^2/Hz$ 
(Fig. \ref{ExpRise}(b) and \ref{ExpRise}(c)), 
while the signal is severely corrupted for 
values $\geq$ 0.04 $mV^2/Hz$ (Figs. \ref{ExpRise}(d) and \ref{ExpRise}(e)).

Fig. \ref{velocity} compares the velocities  of the propagating wavefront as a 
function of the
noise strength for both global and local noise.
The velocity of the propagating wavefront shows an approximate linear increase
with increasing noise strength in both cases. 
Note that the velocity is simply the inverse of the measured arrival times -multiplied
by number of sites.
It is hence well-defined only for low noise levels; in the case of
substantial nucleation of additional thermal kinks-antikink pairs 
this calculated ``velocity'' should be interpreted with caution.
Therefore, the data points in Figs. \ref{velocity} and \ref{Numvelocity} which
correspond to significant noise corruption should be considered as outliers.
For the coupling resistors 
used the velocity for the global noise
case is about 15 \% greater than that seen for local noise.

Besides the earlier onset in the case of global noise, there are also notable 
differences in the mechanism that leads to spurious signals. 
Clearly, for identical elements - unlike spatially uncorrelated noise
 - global noise cannot induce spurious kinks, which then compete with 
the deterministic kink due to the signal. 
The only way of generating a ``false alarm'' at the detector would be 
to phase flip the entire chain, thus requiring a single large fluctuation.
This picture is not fully correct if there are mismatches between the 
elements but illustrates the dominant mechanism of creating the 
signal-masking noise background.
Independent noise sources, however, easily spawn spurious kinks but 
rarely ever cause a global phase flip of the array.
We postulate that the mechanism of noise sustained propagation {\it per se} 
is not very different for local and global noise, presumed the kink width 
is small (i.e. involving a few elements only).
The reason for this conjecture is the local nature of the stochastic escape 
processes that provide for the average effective kink displacement.
As long as the spatial correlation length of the noise is not substantially 
smaller than the kink width, the noise induced ``propulsion '' of the kink 
will be similar for global and local noise.
Note that the findings of K\'ad\'ar {\it et al.} \cite{NEP1} who briefly 
addressed the issue of noise correlation lengths, are consistent with this conjecture.

%%%%%%%%%%%%%%%%%%%%%%%%%%%%%%%%%%%%%%%%%%%%%%%%%%%%%%%%%%%%%%%%%%%%%%%%%%%%%%%%%%%%%%%%%%
\section{Simulations of a model of cardiac tissue}
\label{CardTissue}
Propagation failure of signals due to discreteness of the supporting medium has previously been 
observed in theoretical \cite{Keener} and experimental \cite{myocardiumExp} studies of 
cardiac tissue.
In particular, Keener introduced a modified cable theory, which incorporates 
the discretizing effects 
of the so-called {\it gap junctions} \cite{Keener}.
Gap junctions, characterized by the (relatively high) intercellular resistance $r_g$, provide the 
electrical coupling between cardiac cells. 
Mathematically, the propagation of action potential along cardiac cells is described by various cable theories, which are analogous to wave propagation in one-dimensional conductors (cables). 
Formally, continuous and discrete models describe the wave propagation by either a partial differential equation or - in the latter case - via coupled ordinary differential equations.
{\it Continuous cable theory} either ignores the effects of the gap junctions or replaces the 
cytoplasmic resistance with an effective resistance; in either case the electrical resistance 
is assumed to be spatially homogeneous.
Here, we focus on the opposite assumption that gap junctional resistance is much more important 
than cytoplasmic resistance.
We thus neglect the dynamics within a cell and assume that the propagation of the action 
potential is dominated by the delay caused by the gap junctions \cite{DiscreteAppr}.
Within the context of this {\it discrete cable theory}, we can write the current balance as \cite{Keener}
\begin{equation}
\label{DiscreteModel}
C_m S \frac{d \phi_n}{dt} = \frac{1}{r_g} \left ( \phi_{n+1} - 2 \phi_{n} + \phi_{n-1} \right ) + S I_m (\phi_n)
\end{equation}
where $\phi_n$ is the transmembrane potential for the $n$th cell, $S$ the surface area of cell 
membrane, and $C_m$ is the membrane capacitance per unit area of membrane. 
$I_m$ specifies the inward ionic currents per unit area of membrane and is generally postulated 
to be a function of $\phi_n$ having three zeros.
For simplicity, we choose a simple cubic polynomial $S I_m (\phi) = 12 \sqrt{3} \phi (1-\phi) (\phi - 0.5) + 0.5$ \cite{Keener}.
Note that though similar in appearance, Eqn. (\ref{DiscreteModel}) is {\it not} simply a discretization of its 
continuous analog
\begin{equation}
\label{ContModel}
C_m S \frac{\partial \Phi}{\partial t} = \frac{L^2}{r_g} \frac{\partial^2 \Phi}{\partial x^2}  + S I_m (\phi)
\end{equation}
(where L is the size of the cardiac cell), but stands in its own right as a spatially discrete nonlinear wave
 equation.
The most important observation is that propagation can fail in model (\ref{DiscreteModel}) if $r_g$ is sufficiently 
large, but increasing resistances in Eq. (\ref{ContModel}) can never lead to propagation failure.
Note that $r_g$ depends on the excitability of the tissue.

Fig. \ref{KeenerSpeed} shows a plot of the numerically determined speed (solid line) 
of propagation for model (\ref{DiscreteModel}) as a function of the coupling 
strength $d = 1/r_g$ as well as the analytically obtained 
kink speed $c = \frac{c_0 L}{C_m R_m} 
\sqrt{d}$ (dashed line) for Eq. (\ref{ContModel}).
The reader is referred to Ref. \cite{Keener} for an explanation of $c_0$ and $R_m$.
It is evident that propagation is impossible for $r_g$ larger than a certain critical value $r^*$, which 
turns out to be a monotonic increasing function of excitability \cite{Keener}.

We have performed digital simulations of the stochastic modification of (\ref{DiscreteModel})
\begin{equation}
\label{StochModel}
\frac{d \phi_n}{dt} = \epsilon \left ( \phi_{n+1} - 2 \phi_{n} + \phi_{n-1} \right ) + S I_m (\phi_n) (1+\xi_M(t)) + \xi_A(t)
\end{equation}
using the Euler-Maruyama algorithm \cite{Gard} with a time-step of $dt = 0.05$ and 
coupling strength $\epsilon = 0.07$ (40 elements). 
$\xi_A(t)$ and $\xi_M(t)$ are additive and multiplicative Gaussian white noise, bandlimited in 
practice by the Nyquist frequency $f_N = \frac{1}{2 dt}$. We quantify the noise by its dimensionless variance
 $\sigma^2 = 2 D f_N$, where $2 D$ is the height of the one-sided noise spectrum.

Here, we only consider the case of purely additive noise, $\xi_M(t) = 0$.
By following an analogous procedure as in the experiment we obtain the probabilities 
for successful signal transmission as illustrated in Fig. \ref{NumRise}.
As in the experiment, global noise provides for kink propagation at much 
lower values of $\sigma^2$ than local noise does.
Fig. \ref{Numvelocity} compares the velocities  of the noise propulsed kink as a 
function of $\sigma^2$ for both global and local noise.
The velocity of the propagating wavefront shows an approximately parabolic dependence
on the noise power in both cases. For the coupling strength 
employed global noise leads to speeds about 15 \% greater than that observed for local noise.

%%%%%%%%%%%%%%%%%%%%%%%%%%%%%%%%%%%%%%%%%%%%%%%%%%%%%%%%%%%%%%%%%%%%%%%%%%%%%%%%%%%%%%%%%%%%%%%%%
\section{Detection Criteria} 
\label{DetCriteria}
The decision whether a kink arriving at the last element corresponds to a signal 
injected at the first site constitutes a simple binary hypothesis-testing problem \cite{detection}.
We assign the null hypothesis $H_0$ to ``no signal injected'' and the opposite for 
the alternative hypothesis $H_1$. Denote the according decisions $D_i$ as the judgment 
that hypothesis $H_i$ was in effect.
Clearly, there are two possible errors: A so-called {\it Type I error} occurs when 
making the decision that $H_1$ was in effect while the contrary is true. Borrowing 
notation from radar detection, we refer to this as the probability of 
{\it false alarm} $P_f = P(D_1 | H_0)$.
On the other hand, if $H_1$ was in effect to generate the data, and we decide $D_0$, 
then we have committed a {\it Type II error}, which in radar is referred to as a 
{\it missed detection} \cite{detection}.
We do that with some probability $P(D_0 | H_1)$ which is related to the 
{\it probability of detection} $P_d = P(D_1 | H_1)$ in an inverse fashion: $P_d  = 1 - P(D_0 | H_1)$.

In our experiment, there is no uniquely defined, objectively ``best'' noise level without 
first defining a decision strategy.
Two suitable approaches are (i) the {\it Neyman-Pearson strategy}, in which the probability 
of detection $P_d$ is maximized while specifying an upper bound for the false alarm probability 
$P_f$, and (ii) {\it Bayes' rule} which assigns costs to 
the various outcomes of the decision process, such as correctly detecting a signal or being 
``deceived'' by a spurious kink.
The optimum noise level in the latter case would be the one, which minimizes the total average cost. 

\noindent In communication systems one is usually interested in the total probability of error 
$P_e = P(D_1 | H_0) + P(D_0 | H_1) = P_f + 1 - P_d$.
Though in the experimental setup there is no inherent time-scale, i.e. the decision when to 
reset the chain is rather arbitrary, in digital communication applications we would expect 
information bits to be sent at a constant rate.
Hence, we choose a reasonable time interval at the end of which we measure the probabilities 
of false alarm $P_f$ and missed detection $1-P_d$ as a function of noise power.
Then, $P_f$ is simply the value of the dashed line in Fig. \ref{NumRise} at time = 300, and $P_d$ 
is the corresponding value for the solid line.
For local and global noise, the total probability of error $P_e$ is displayed in Fig. \ref{NumPe}; 
for very low and rather high values of the noise power, $P_e$ is almost one. 
However, there exists an optimal noise strength for which both $P_f$ and $1-P_d$ nearly vanish, 
resulting in a sharp minimum of $P_e$.

%%%%%%%%%%%%%%%%%%%%%%%%%%%%%%%%%%%%%%%%%%%%%%%%%%%%%%%%%%%%%%%%%%%%%%%%%%%%%%%%%%%%%%%%
\section{Theory}
\label{Theory}
 The results of Sects. \ref{ExpResults} - \ref{DetCriteria} touch upon two central properties of kink
statistics in a discrete bistable chain, namely the Brownian motion of an
individual kink and the nucleation of kink-antikink pairs in the
presence of an external static bias (or d.c. forcing term).
In the absence of a precise model for the 2-state potential that
describes the phase shifts in the diode resonators of Sec. \ref{ExpResults} or drives
the transmembrane potential of Sec. \ref{CardTissue}, we speculate that the {\it in
situ} bistability of our arrays can be rendered satisfactorily in terms of
a Double
Quadratic (DQ) potential, that is by two parabolas
displaced by a distance $2a$ (see Fig.
\ref{DQ}a). A discrete DQ model is likely to capture, at least
qualitatively, the essential features of the array dynamics investigated
above, while affording substantial simplifications in its analytical
treatment. The analysis of this Section can be carried over, with more
mathematical effort, to the $\phi^4$ model of Sec. \ref{CardTissue}, as well.
 
The DQ model has been studied both in the continuum \cite{currie} and in  
the discrete case \cite{willis}. In dimensionless units the DQ
Hamiltonian reads
\begin{equation}
\label{V1}
 {H\over {H_0}} =l \sum_n \left \{ {{\dot \phi_n^2} \over 2}+{{c_0^2}
\over {4l^2}}
 [(\phi_n-\phi_{n-1})^2 +(\phi_n-\phi_{n+1})^2]+
 {{\omega_0^2} \over 2}(|\phi|-1)^2
 \right \}, 
\end{equation}
with $H_0=ma^2/l$.  Each $\phi_n$ can be regarded as the 
displacement (in units of $2a$)
of the n-th chain site
with mass  $m$, $c_0$ and $\omega_0$ represent
respectively the limiting speed
and frequency of the phonon modes propagating along the chain and $l$    
denotes the chain lattice constant ( for instance, in Eqn.
(\ref{DiscreteModel}) $l$ was
set to one). The ratio $c_0^2 /l^2$, which quantifies the effectiveness  
of the coupling between two adjacent bistable units, is the coupling
constant of our model. The importance of the discreteness effects is
measured by the {\it discreteness} parameter 
\begin{equation}
\label{V2}
\gamma = {{c_0^2}\over{\omega_0 l}} \equiv {d \over l},
\end{equation}
namely the ratio of the kink length $d$ to the chain spacing $l$.

\subsection{The continuum limit}
\label{TheoryA}
In the continuum (or displacive) limit $\gamma \rightarrow \infty$ the
Hamiltonian (\ref{V1}) can be expressed as the line integral of the
Hamiltonian density
\begin{equation}
\label{V3}
H[\phi] = {{\phi_t^2}\over 2} + c_0^2{{\phi_x^2}\over 2} +V[\phi],
\end{equation}  
where the string field $\phi(x,t)$ is defined as $\lim_{l\rightarrow 0}
\phi_{x/l}(t)$ and $V[\phi]={{\omega_0^2}\over 2}(|\phi|-1)^2$.
The statistical mechanics of the continuum DQ model can be worked out
analytically in great detail \cite{currie}. In particular, we know that
the kink ($\phi_+$) and the antikink solutions ($\phi_-$)
\begin{equation}
\label{V4}
\phi_{\pm}(x,t)=\pm \mbox{sgn}[x-X(t)][1-\exp^{-{|x-X(t)|}\over
{d\sqrt{1-u^2/c_0^2}}}]
\end{equation}
can be regarded as relativistic quasi-particles with size
$d=c_0/\omega_0$, mass $M_0=E_0/c_0^2=1/d$ (or rest energy $E_0=\omega_0
c_0$) and center of mass $X(t)=X_0+ut$.

At low temperatures, $kT \ll E_0$, any string configuration can be
represented as a linear superposition of randomly distributed kinks and
antikinks floating on a phonon bath. A DQ string
in equilibrium at temperature $T$ and with boundary conditions
$\phi(-\infty,t) =\phi(+\infty,t)$ bears naturally a dilute gas of thermal
kink-antikink pairs with density
\begin{equation}
\label{V5}
n_0(T)= {1 \over {2\sqrt{2}}} {1\over d} \left ({{E_0}\over {kT}} \right)^{1/2}
e^{-E_0/kT}.
\end{equation}
The qualification thermal underscores the fact that $n_0$ pairs per
unit of length (with $n_0d \ll 1$) are
being generated by thermal fluctuations alone, irrespective of any
geometric constraint at the boundaries (see discussion of Fig.
\ref{ExpKinkVel} in Sec. \ref{ExpResults}).

\subsection{Kink Brownian motion}
\label{TheoryB}
The $\phi_{\pm}(x,t)$ solutions (\ref{V4}) tend to travel with arbitrary
constant speed $u < c_0$, unless perturbed by the coupling to a heat bath
or by an external field of force. The simplest heat-bath model was
obtained \cite{langevin} by adding
a viscous term $-\alpha \phi_t$ and a zero-mean, Gaussian {\it local}
noise source $\zeta(x,t)$ to the string equation of motion corresponding
to the Hamiltonian density (\ref{V3}), that is
\begin{equation}
\label{V6}
\phi_{tt} -c_0^2 \phi_{xx} -\omega_0^2 \mbox{sgn}[\phi](|\phi|-1)=-\alpha
\phi_t
+F +\zeta(x,t).
\end{equation}
Note that all our experiments and simulations have been carried out in the
overdamped limit, $\alpha \gg \omega_0$, and in the presence of an
additional
sub-threshold force $F$ with $F<\omega_0^2$, also incorporated in Eq.
(\ref{V6}).
Thermalization is imposed here 
by choosing the noise autocorrelation function $\langle \zeta(x,t)
\zeta(x',t')
\rangle = 2 \alpha kT \delta (t-t') \delta(x-x')$.

A single kink (antikink) subjected to thermal fluctuations undergoes
driven Brownian motion with Langevin equation
\begin{equation}
\label{V7}
\dot X = \mp {{2F}\over {\alpha M_0}} + \eta(t), 
\end{equation}
where $\eta(t)$ is a zero-mean-valued Gaussian noise with strength $D=
kT/\alpha M_0$ and autocorrelation function $\langle \eta(t) \eta(0)\rangle
=2D\delta(t)$. As apparent from Eq. (\ref{V7}), the external bias pulls
$\phi_{\pm}$ in opposite
directions with average speed $\pm u_F$ and $u_F=  2F/\alpha M_0$.

If the local fluctuations are spatially correlated, say 
\begin{equation}
\label{V8}
\langle \zeta(x,t) \zeta(x',t') \rangle = 2 \alpha kT \delta(t-t')
[e^{-|x-x'|/\lambda}/2\lambda],
\end{equation}
the noise strength $D$  changes into \cite{langevin}
\begin{equation}
\label{V9}
D(\lambda)={{D d} \over {\lambda + d}}\left ( 1 + {{\lambda} \over
{\lambda + d}} \right ).
\end{equation}
As speculated in Sec. \ref{ExpResults}, for noise correlation length $\lambda$ smaller
than the kink size $d$ possible spatial dishomogeneities become
negligible, i.e. $D(\lambda) \simeq D$ for $\lambda \ll d$
\cite{disorder}. The {\it
global} noise regime simulated both experimentally in Sec. \ref{ExpResults} and
numerically in Sec. \ref{CardTissue} corresponds to the limit $\lambda \rightarrow
\infty$ of the source $\zeta(x,t)$ rescaled by the normalization factor
$\sqrt{2\lambda}$; the Langevin equation (\ref{V7}) still applies, but the
relevant noise strength is now $\lim_{\lambda \rightarrow \infty} 2\lambda
D(\lambda)=4D$. This accounts for the observation that global noise
sustains kink propagation more effectively than local noise. Note that the
enhancement factor of 4, more exactly $4 a$, is nothing but twice the distance of the DQ
potential minima (in dimensionless units).

Another important property of global noise is that it cannot trigger the
nucleation of a kink-antikink pair and, therefore, minimizes the chances
of a "false alarm" (see Fig. \ref{ExpRise}). For this to occur it would be
necessary that a spatial deformation of a stable string configuration
(vacuum state) be generated large enough for the external bias to succeed
in making it grow indefinitely. Such a 2-body nucleation mechanism would
require a {\it local} breach of the $\phi \rightarrow -\phi$ symmetry of
the DQ equation (\ref{V6}), which can be best afforded in the presence of
uncorrelated {\it in situ} fluctuations \cite{nucleation}.

The nucleation rate, namely the number of kink-antikink pairs generated
per unit of time and unit of length, can be easily computed by combining
the nucleation theory of Ref. \cite{nucleation} with the analytical
results of Ref. \cite{currie} for the DQ theory. For values of the string
parameters relevant to  Secs. \ref{ExpResults} - \ref{DetCriteria}, that is for $kT$ and  $Fd \ll E_0$ 
the stationary DQ nucleation rate can be approximated to \cite{nucleation}
\begin{equation}
\label{V10}
\Gamma_1(T)={{2n_0(T)} \over {\tau(T)}} = 2u_Fn_0^2(T),
\end{equation}
if $Fd \ll kT$, or $\Gamma_2(T)={1 \over 2}\sqrt{{{kT} \over {Fd}}}
\Gamma_1(T)$, if $kT \ll Fd \ll E_0$ \cite{Fd}. 
For an overdamped string, $\alpha \gg \omega_0$ the time constant
$\tau(T)$ amounts to the kink (antikink) lifetime prior to a destructive
collision with an antikink (kink). Both estimates for the DQ nucleation
rate clearly show that spontaneous nucleation of thermal pairs may degrade
appreciably local-noise sustained propagation of injected (or geometric)
kinks only for thermal energy fluctuations of the order of the kink rest
energy.

\subsection{The Peierls-Nabarro potential}
\label{TheoryC}
Let us go back now to the case of a discrete DQ chain. Discreteness (with
parameter $\gamma$) affects the kink dynamics on two accounts:

(i) The profile of a {\it static} kink (antikink) $\phi_{\pm}(x,0)$ is
deformed into \cite{willis}
\begin{equation}
\label{V11}
\phi_{\pm,n}^{(s)} = \pm \mbox{sgn}[n-N] [1-Z_{\nu} \nu^{|n-N|}],
\end{equation}
with $Z_{\nu}=2\sqrt{\nu}/1+\nu$, $N=m+1/2$, $m=0, \pm 1, \pm2, \dots$,
and $\nu=[\sqrt{1+4\gamma^2}-1]/[\sqrt{1+4\gamma^2}+1]$. To make contact
with the displacive solution $\phi_{\pm}(x,0)$ one must replace $nl$ with
$x$, $Nl$ with $X_0$ and take the continuum limit $\gamma \rightarrow
\infty$ (so that $\nu \simeq 1-1/\gamma$). Note that the spatial extension
of the discrete kink solutions  $\phi_{\pm,n}^{(s)}$ increases
monotonically
with $\gamma$. As $\gamma$ decreases below unity,  $\phi_{\pm,n}^{(s)}$
approaches a step function (order-disorder limit);

(ii)  $\phi_{\pm,n}^{(s)}$ is centered midway between two chain sites due
to the confining action of an effective [or Peierls-Nabarro (PN)]
potential \cite{willis}. The PN potential describes the spatial modulation
of the  $\phi_{\pm,n}^{(s)}$ rest energy as its center of mass is
moved across one chain unit cell, say from $ml$ up to $(m+1)l$.

As a result, according to the Langevin equation approach of Sec. \ref{TheoryB}, 
the $\phi_{\pm,n}^{(s)}$ center of mass $X(t)$ diffuses on a periodic,
piece-wise harmonic potential with constant $l$ and angular frequency
$\omega_{PN}$, that is \cite{willis}
\begin{equation}
\label{V12}
\alpha \dot X = -\omega^2_{PN}[X-l(\mbox{Int}[X/l] -1/2)] \mp 2F/M_0
+\alpha
\eta(t), 
\end{equation}
where $\omega^2_{PN} \simeq (1+\nu)\omega_0^2$ and Int$[X/l]$ denotes the
integer part of $X$ in units of $l$. Note that $\omega_{PN} \rightarrow
\omega_0$ and $\omega_{PN} \rightarrow
\sqrt{2} \omega_0$ in the highly discrete and continuum limit,
respectively. The energy barriers of the PN potential are thus (almost)
quadratic in $l$.

The one-dimensional Langevin equation (\ref{V12}) has been studied in
great
detail by Risken and coworkers \cite{risken}. In the noiseless limit,
$\eta(t) \equiv 0$, the process $X(t)$ is to be found either in a locked
state with $\langle \dot X \rangle=0$, for
$4F/M_0 <\omega_{PN}^2$ , or in a running
state with  $\langle \dot X \rangle \simeq u_F$, for 
$4F/M_0 >\omega_{PN}^2$. This is indeed the depinning (or
locked-to-running) transition described in Fig. \ref{ExpKinkVel}.
At {\it finite} temperature the stationary velocity  $\langle \dot X
\rangle=u(T)$ can be cast in the form following,
\begin{equation}
\label{v13}
{{u(T)} \over {u_F}} = {1 \over {\delta}} {{1 - e^{-\delta}} \over
{A - B(1-e^{-\delta})}},
\end{equation}
where $\delta= 2 Fl/kT$ and the quantities $A$ and $B$ can be computed
numerically with minimum effort \cite{formula}.
The ratio $u(T)/u_F$ is the rescaled
 $\phi_{\pm,n}^{(s)}$  mobility; it crosses from 0 (locked state) over
to 1 (running state) continuously in a relatively narrow neighborhood
of the threshold value $F_{th}=M_0\omega_{PN}^2/4$; moreover, $u(T)/u_F$
increases monotonically with $T$ at fixed bias. Such a temperature
dependence of the kink mobility explains the sequences of rise curves
in Figs. \ref
{ExpRise} and \ref{NumRise}, where kink propagation seems to speed up 
on raising the noise level.

%%%%%%%%%%%%%%%%%%%%%%%%%%%%%%%%%%%%%%%%%%%%%%%%%%%%%%%%%%%%%%%%%%%%%%%%%%%%%%
\section{Summary}
In conclusion, the present analysis confirms our speculation that
 the apparent SR behavior of the efficiency of noise-sustained
transmission of kink-like signals along
a bistable chain results from two competing mechanisms,
both controlled by noise:
The driven diffusion dynamics of stable noninteracting kinks, 
which increases exponentially  with the temperature
in the vicinity of the depinning transition
(propulsion mechanism); The detection of spurious signals,
as thermal kink-antikink pairs nucleate with exponentially
increasing rates, thus corrupting the propagated signal
(garbling mechanism).

If the spatial distribution of the noise was constrained to a small neighborhood around 
the kink and zero along the rest of the chain, fast and efficient noise supported signal 
transmission without false alarms would be realizable. 
This seemingly artificially constructed scenario can be achieved naturally by considering the 
case of purely multiplicative noise \cite{multNoise}.
A detailed study of noise sustained propagation in the presence of multiplicative
fluctuations is beyond the scope of this work.

During the preparation of this manuscript the authors learned about recent
results on propagation failure in the context of cell differentiation \cite{CellDiff}.
Utilizing a highly simplified model composed of coupled bistable elements, 
the authors furnish evidence for the discrete nature of chemical signaling waves 
propagating through a chain of cells.
We speculate that fluctuations, inherent in biological systems, might 
play a significant role in the details of cell differentiation processes.

We acknowledge the Office of Naval Research for financial support.
ML, NC and EH warmly thank D. Cigna for very significant contributions to the 
experimental setup.
%%%%%%%%%%%%%%%%%%%%%%%%%%%%%%%%%%%%%%%%%%%%%%%%%%%%%%%%%%%%%%%%%%%%%%

%%%%%% FIG 1%%%%%%%%%%%%%%%%%%
\begin{figure} 
\begin{center}
\leavevmode
\end{center}
\caption{Experimental circuit arrangement for any node (i,j). 
We either add noise locally, as shown, or globally by adding 
one noise source to the main drive. Coupling is provided by the resistors $R_C$.}
\label{circuit}
\end{figure}

%%%%%% FIG 2%%%%%%%%%%%%%%%%%%
\begin{figure}
\begin{center}
\leavevmode
\end{center}
\caption[Kink speed as function of bias]
{Kink velocity as a function of the bias for an intermediate value of the coupling resistor $R_C$. The kink speed  shows an an approximate linear
decrease with bias down to a cut off value of approximately 0.9 units.}
\label{ExpKinkVel}
\end{figure}

%%%%%% FIG 3%%%%%%%%%%%%%%%%%%
\begin{figure}[t]
\begin{center}
\leavevmode
\end{center}
\caption[4 Rise Curves with and without forcing, LOCAL AND GLOBAL NOISE]
{Arrival-probablity  of a signal at the end of the chain with (solid line) and 
without (dashed curves) induced kink at the beginning, for five local and 
global noise levels increasing from top to bottom.}
\label{ExpRise}
\end{figure}

%%%%%% FIG 4%%%%%%%%%%%%%%%%%%
\begin{figure}[t]
\begin{center}
\leavevmode
\end{center}
\caption[Average kink velocity]{Experimentally observed 
average kink velocities as a function of noise strength for both global and local noise.
Note that the last (two) data point(s) for local (global) noise are corrupted by noise and
should be interpreted as velocity only with caution.}
\label{velocity}
\end{figure}

%%%%%% FIG 5%%%%%%%%%%%%%%%%%%
\begin{figure}[t]
\begin{center}
\epsfxsize=2.5in
\leavevmode
%\epsffile{Fig5.eps}
\end{center}
\begin{center}
\begin{minipage}{3in}
\caption[Kink speed]{{Kink speed as a function of coupling strength 
$d = \frac{1}{r_g}$ for the continuous model Eqn. (\ref{ContModel}) 
(dashed line) and the discrete model Eqn. (\ref{DiscreteModel}) (solid line). 
Note that propagation in the 
discrete model is impossible for $r_g > r^*$ ($d < d^* = \frac{1}{r^*}$). 
For this simulation, $C_m S = 1$ and 
$S I_m (\phi) = 12 \sqrt{3} \phi (1-\phi) (\phi - 0.5) + 0.5$}
\label{KeenerSpeed}}
\end{minipage}
\end{center}
\end{figure}

%%%%%% FIG 6%%%%%%%%%%%%%%%%%%
\begin{figure}[t]
\begin{center}
\leavevmode
\end{center}
\caption[4 Rise Curves with and without forcing, LOCAL AND GLOBAL NOISE,SIMULATIONS]
{Arrival probabilities for the discrete model Eqn. (\ref{DiscreteModel}) with (solid line) 
and without (dashed curves) induced kink at site 1, for five local and 
global noise levels increasing from top to bottom. 
The propagation distance spans 40 elements. Note that we employ dimensionless units
for the noise power $\sigma^2$.}
\label{NumRise}
\end{figure}

%%%%%% FIG 7%%%%%%%%%%%%%%%%%%
\begin{figure}[t]
\begin{center}
\leavevmode
\end{center}
\caption[Average kink velocity]
{Numerically measured 
average kink velocities as a function of noise strength $\sigma^2$ (dimensionless)for both global and local noise.
It is insightful to compare these values with Fig. \ref{KeenerSpeed}.
As in Fig \ref{velocity}, the last data point includes a large fraction of spurious kinks
which results in an artificially high value for the velocity.}
\label{Numvelocity}
\end{figure}

%%%%%% FIG 8%%%%%%%%%%%%%%%%%%
\begin{figure}[t]
\begin{center}
\leavevmode
\end{center}
\caption[Pe for simulations as a function of noise]
{Total probability of error $P_e$ as a function of noise variance $\sigma^2$ (dimensionless)
for global (solid line) and local noise (dashed line). For a range of optimal 
noise strengths $P_e$ virtually vanishes. The qualitative behavior is robust to variations 
in the measurement time, which here is taken to be 300 time units.}
\label{NumPe}
\end{figure}

%%%%%% FIG DQ%%%%%%%%%%%%%%%%%%
\begin{figure}[t]
\begin{center}
\epsfxsize=4.5in
\leavevmode
%\epsffile{Fig9.eps}
\end{center}
\begin{center}
\begin{minipage}{3in}
\caption[DQ]
{{The Double Quartic (DQ) model for $\omega_0=c_0=\alpha=1$ and discreteness
parameter $\gamma=1$. (a) The DQ potential $V[\phi]$ of Eqs. (\ref{V1}) and (\ref{V3});
(b) The PN potential $V(x,F)$ of Eq. (\ref{V12}) and \cite{formula} for
$F/F_{th}=0.08$; (c) Kink stationary velocity $u(T)$ versus bias intensity
$F$ for $\omega_0^2/kT=20$ (curve 1) and 50 (curve 2). At $T=0+$ the
limiting curve (dashed) is given by $u/u_F=0$ for $F<F_{th}$ and $u/u_F=
\{(F/F_{th})\ln
[F/(F-F_{th})]\}^{-1}$ for $F>F_{th}$. All quantities plotted are dimensionless.}
\label{DQ}}
\end{minipage}
\end{center}
\end{figure}


\begin{references}
\bibitem{DiscreteAppr} Of course, this is also an approximation. 
In future work we will replace the discrete cable theory with the 
{\it modified cable theory} \cite{Keener} which incorporates both the discretizing 
nature of the gap junctions as well as the continuous propagation within a cell.
\bibitem{Fd} $F d$ is essentially the work done by moving the kink through a distance of
the order of its radius $d$. The inequality $Fd \gg kT$ identifies the kink as a pointlike 
object in comparison with thermal fluctuations.
\bibitem{formula} The derivation of the (exact !) expression Eqn. (\ref{v13})
follows Chap. 11 in \cite{risken}.
The quantities $A$ and $B$ are defined as follows:
$A=\int_0^le^{-V(x,F)/kT}dx~\int_0^le^{V(x',F)/kT}dx'$ and
$B=\int_0^le^{-V(x,F)/kT}dx~\int_0^xe^{V(x',F)/kT}dx'$, with $V(x,F)$
denoting the tilted PN potential ${1\over 2}M_0\omega_{PN}^2(x-{l \over
2})^2 -2Fx$.
\end{references}
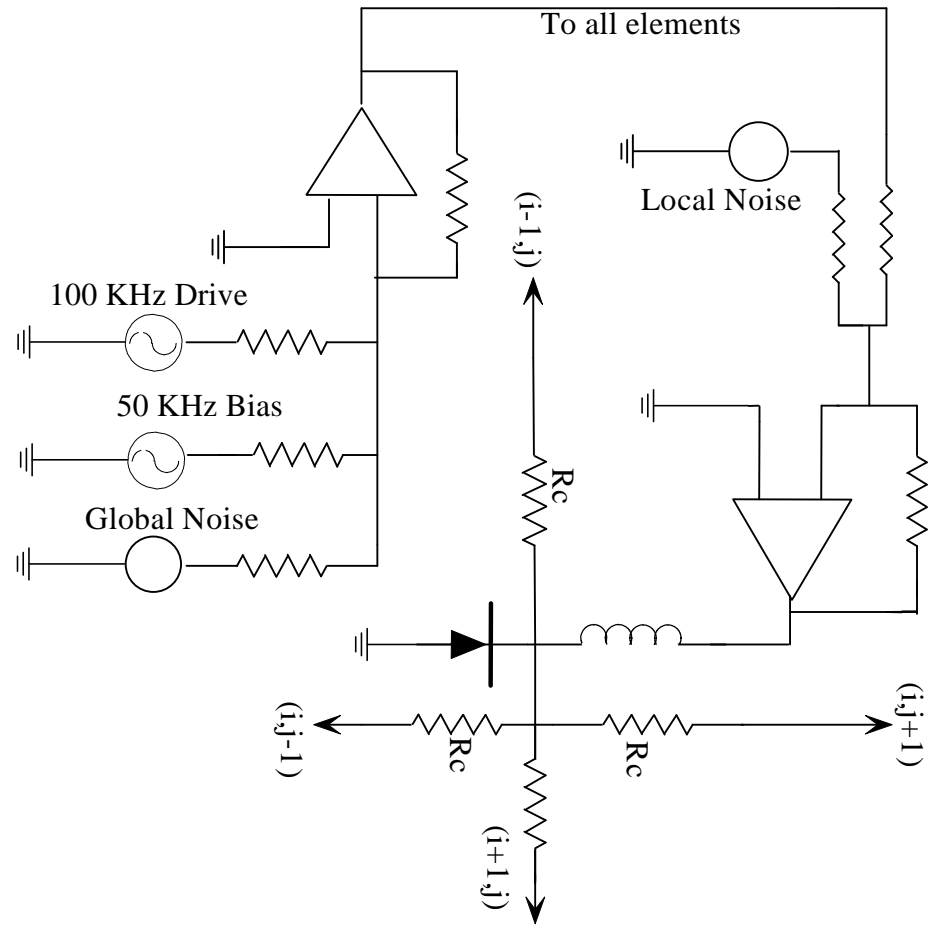
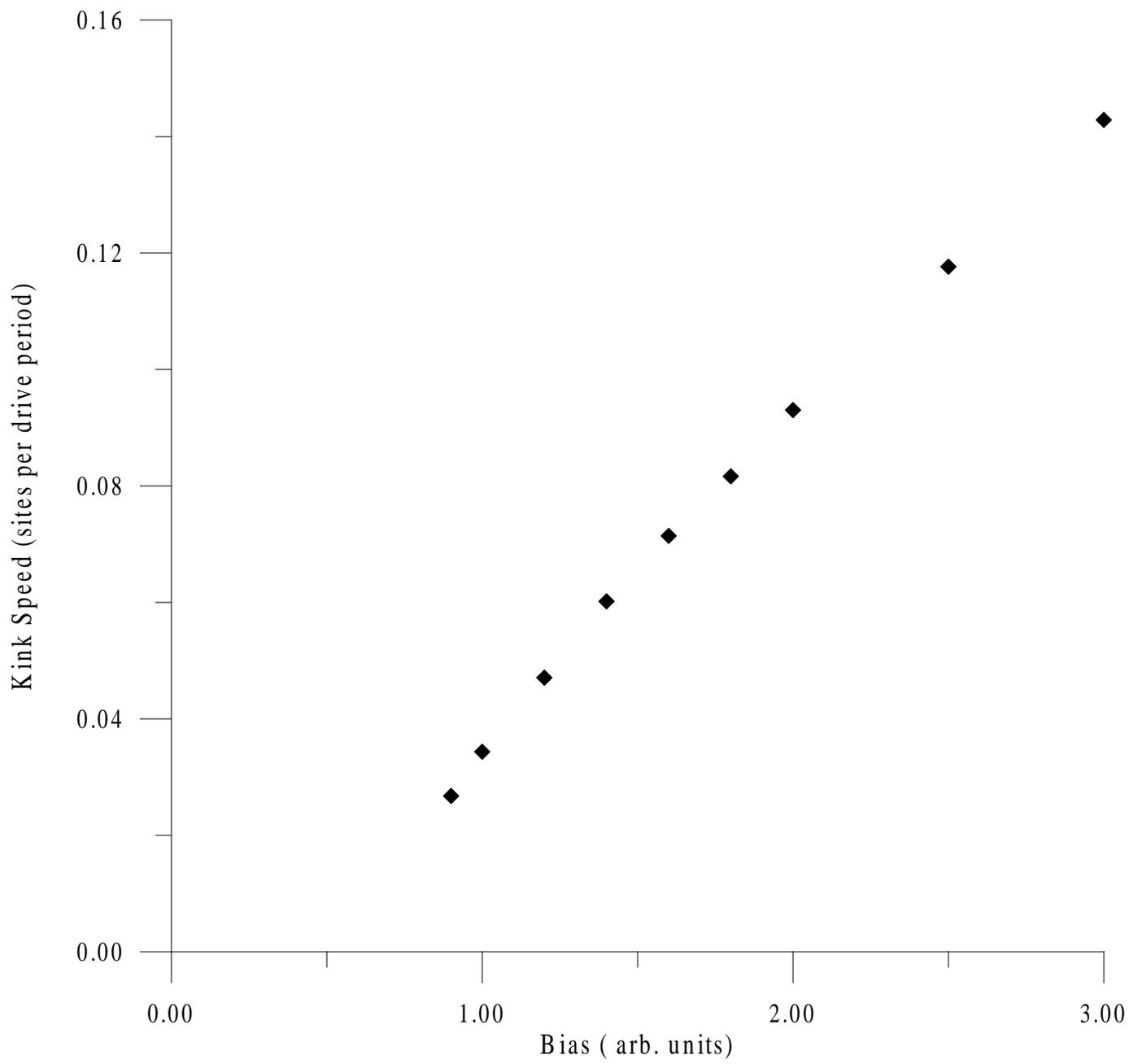
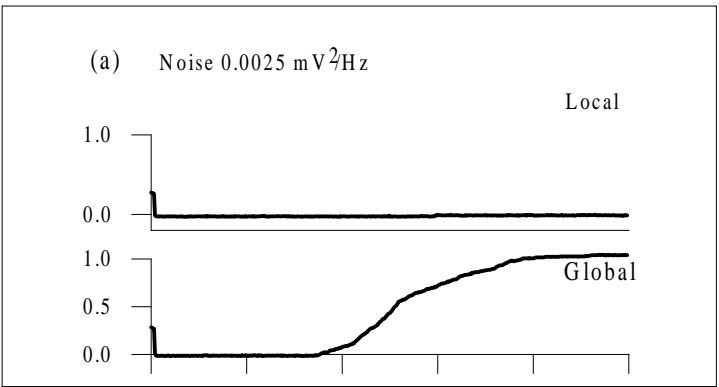
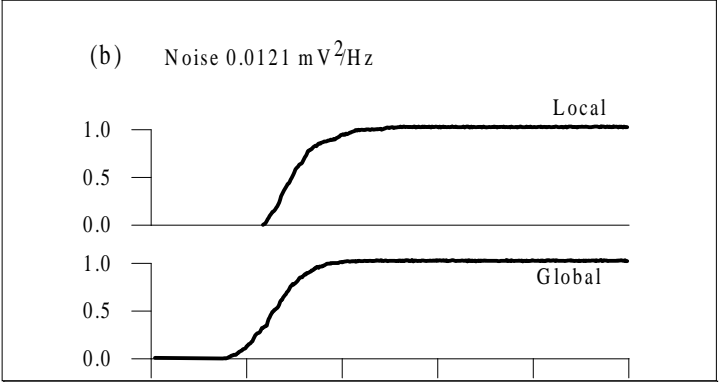
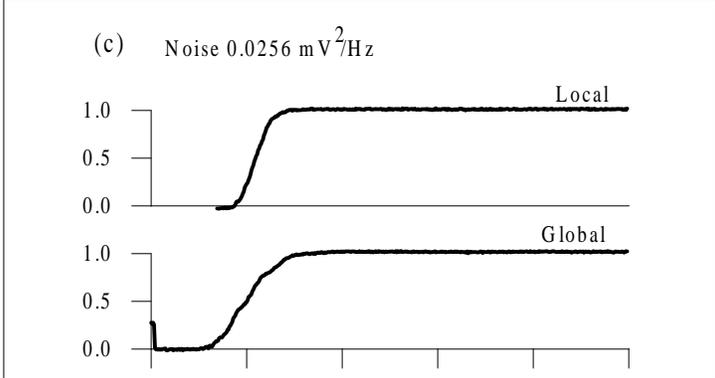
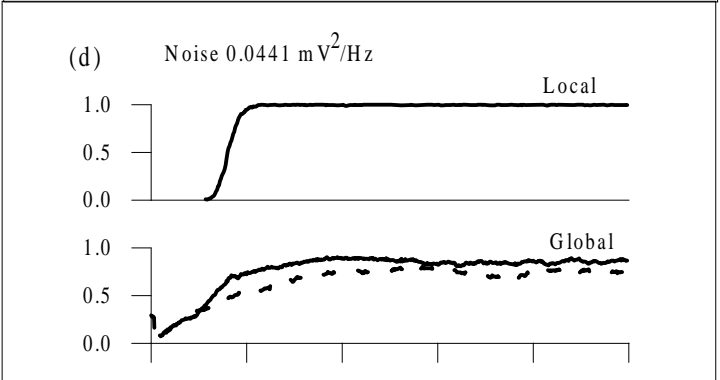
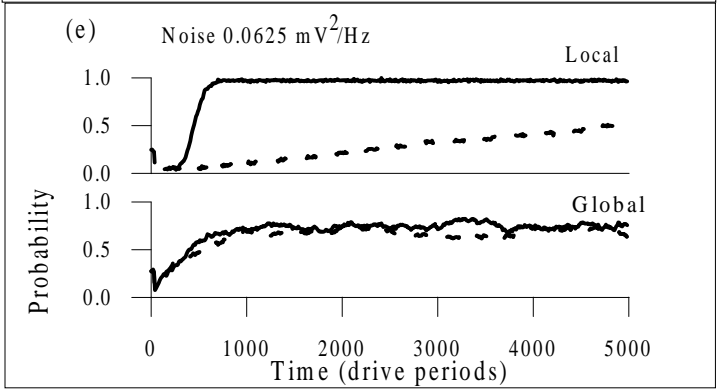
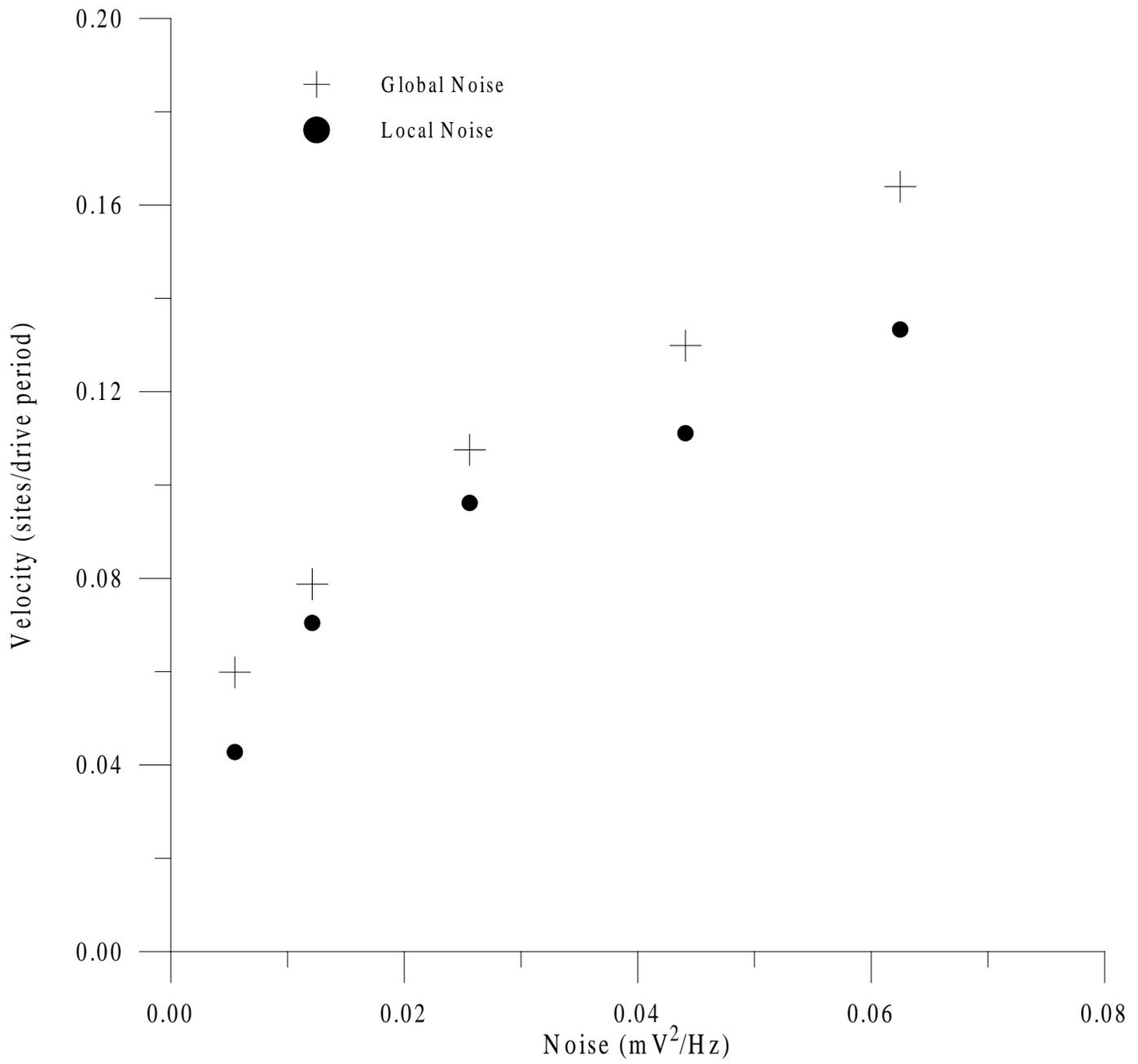
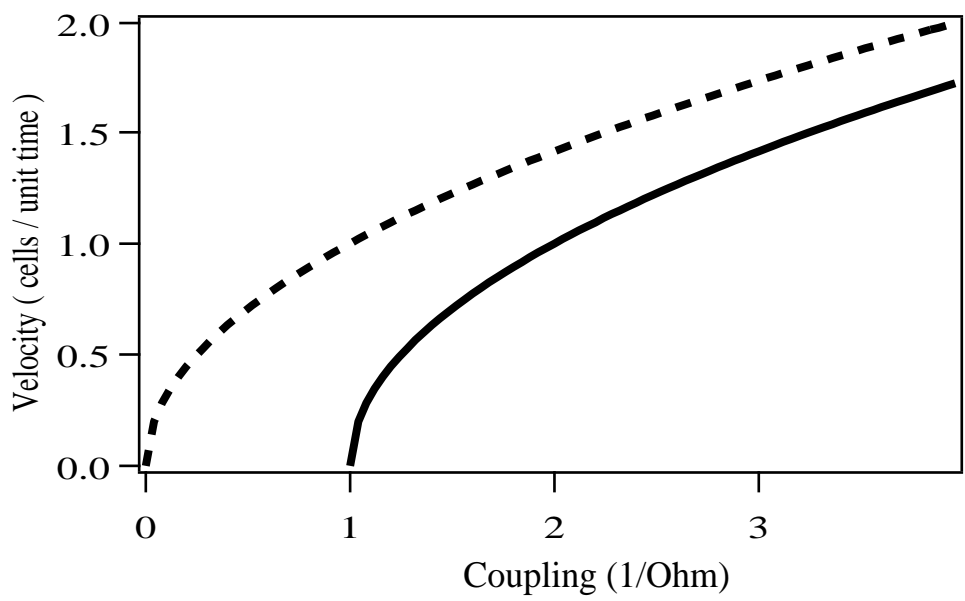
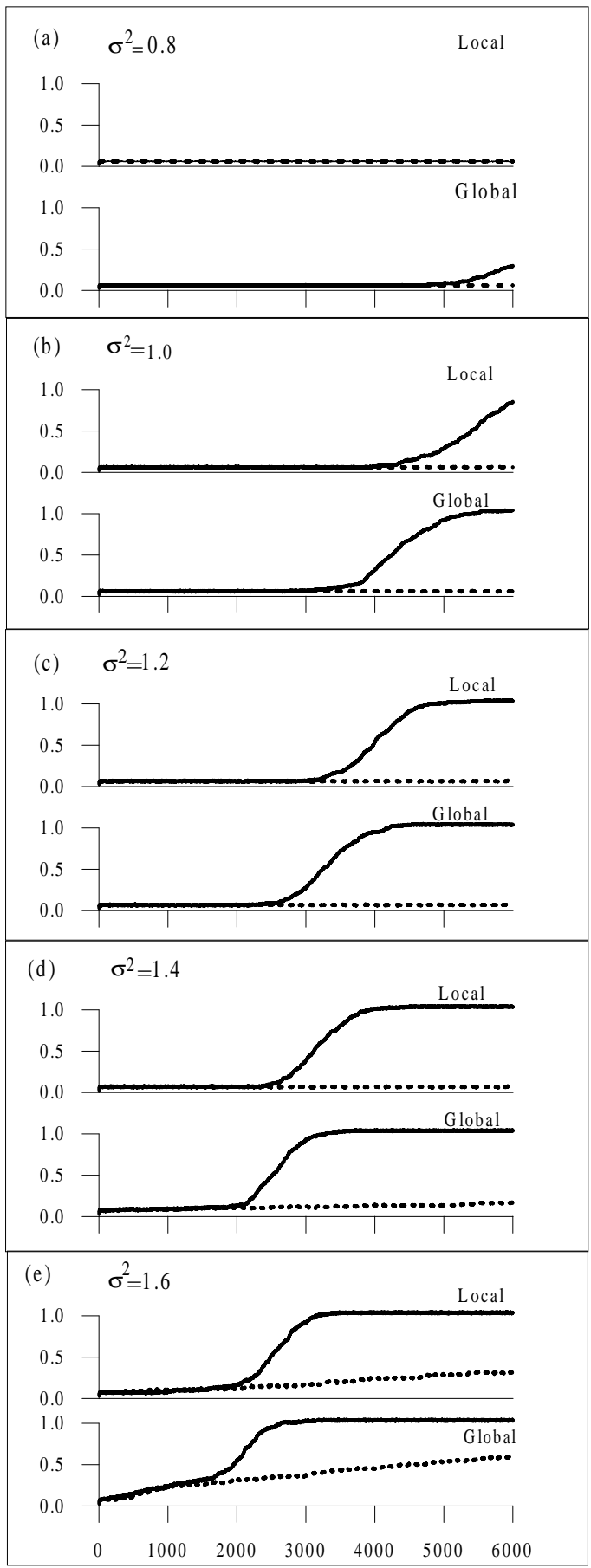
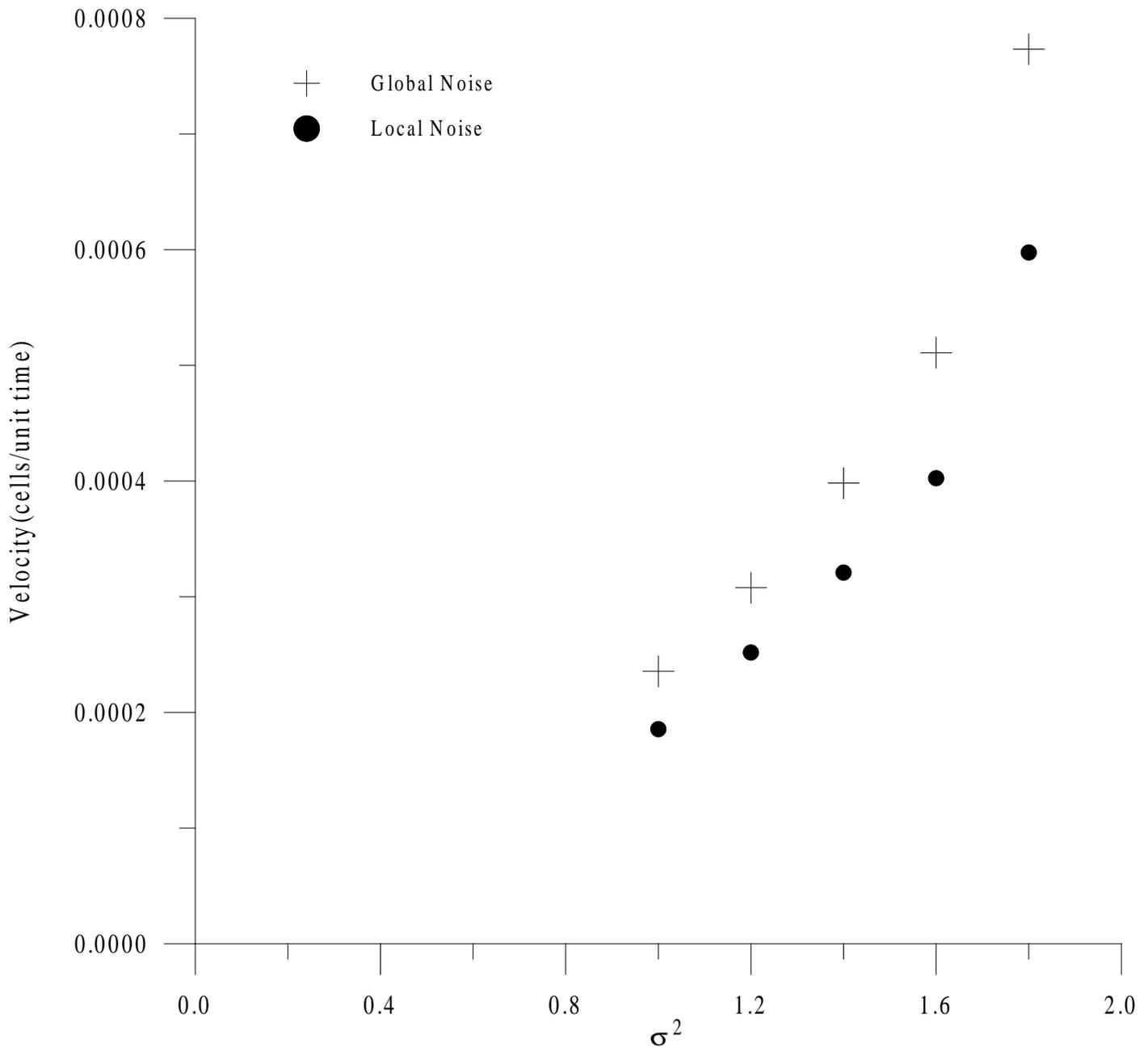
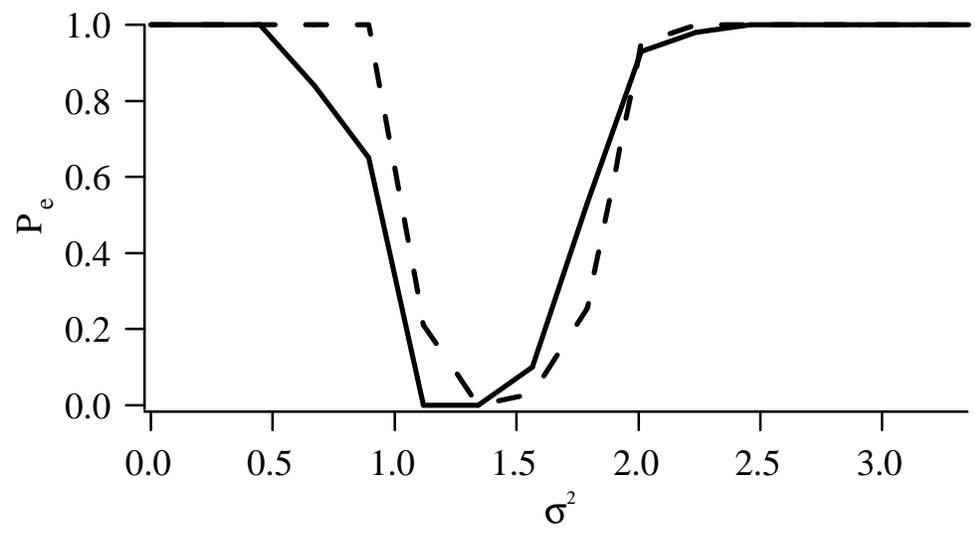
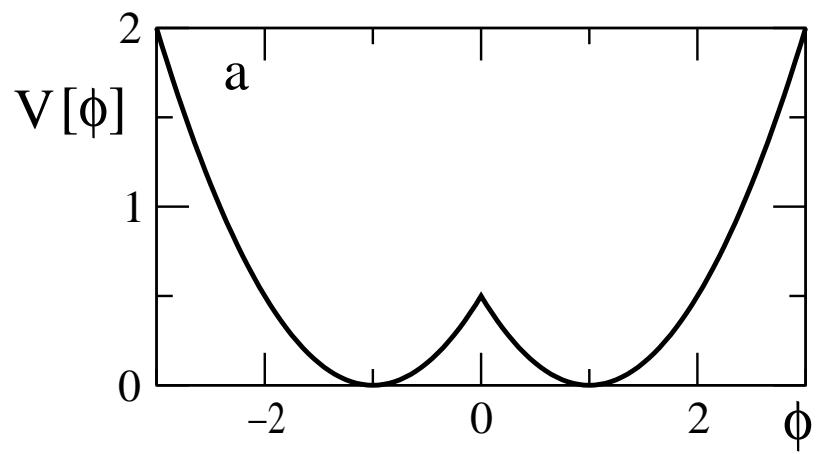
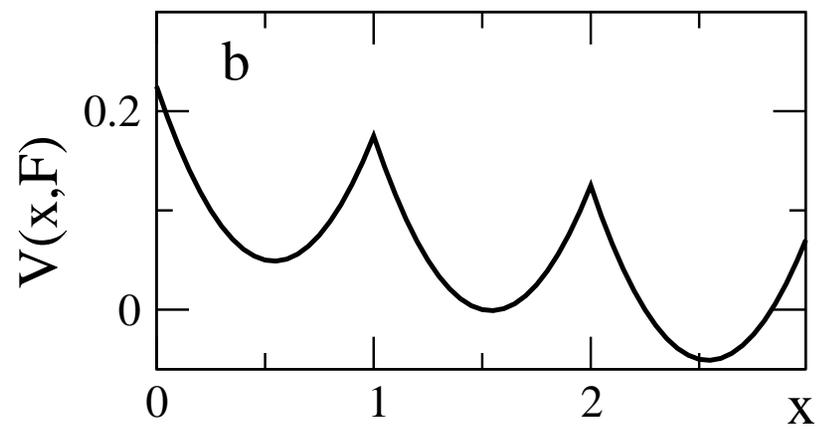
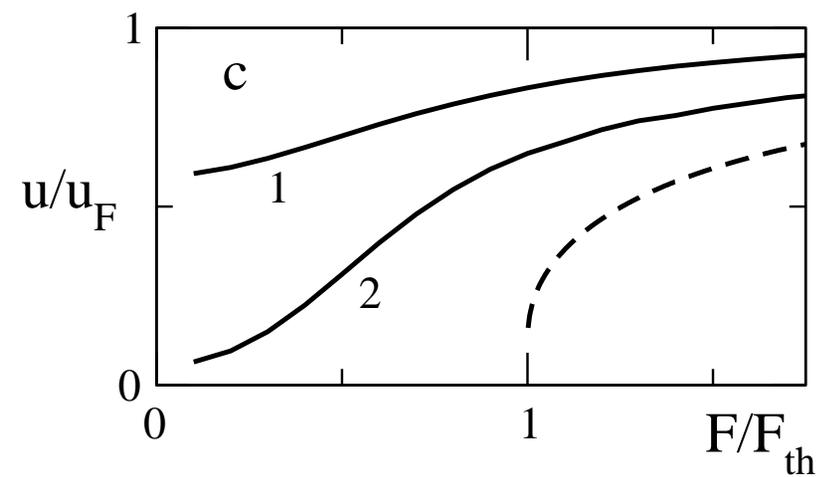
\end{document}